\documentclass[12pt,preprint]{aastex}

\shorttitle{Leo\,A. Suprime-Cam Stellar Photometry}
\shortauthors{R. Stonkut\.{e} et al.}

\begin{document}

\title{Dwarf Irregular Galaxy Leo\,A. \\ Suprime-Cam Wide-Field Stellar Photometry \altaffilmark{1}}

\author{Rima Stonkut\.{e} \altaffilmark{2}, Nobuo Arimoto \altaffilmark{3,4}, Takashi Hasegawa \altaffilmark{5}, Donatas Narbutis \altaffilmark{2}, Naoyuki Tamura \altaffilmark{6}, Vladas Vansevi\v{c}ius \altaffilmark{2,7}}

\altaffiltext{1}{Based on data collected at the Subaru Telescope, which is operated by the National Astronomical Observatory of Japan}
\altaffiltext{2}{Center for Physical Sciences and Technology, Savanori\c{u} 231, Vilnius LT-02300, Lithuania}
\altaffiltext{3}{Subaru Telescope, National Astronomical Observatory of Japan, 650 North A'ohoku Place, Hilo, HI 69720, USA}
\altaffiltext{4}{Department of Astronomical Science, The Graduate University of Advanced Studies, Mitaka, Tokyo 181-8588, Japan}
\altaffiltext{5}{Gunma Astronomical Observatory, 6860-86 Nakayama, Takayama-mura, Agatsuma-gun, Gunma 377-0702, Japan}
\altaffiltext{6}{Kavli Institute for the Physics and Mathematics of the Universe (WPI), Todai Institutes for Advanced Study, the University of Tokyo, 5-1-5 Kashiwanoha, Kashiwa, Chiba 277-8583, Japan}
\altaffiltext{7}{Vilnius University Observatory, \v{C}iurlionio 29, Vilnius LT-03100, Lithuania; vladas.vansevicius@ff.vu.lt}

\begin{abstract}
We have surveyed a complete extent of Leo\,A -- an apparently isolated gas-rich low-mass dwarf irregular galaxy in the Local Group. The $B$, $V$, and $I$ passband CCD images (typical seeing $\sim$0.8\arcsec) were obtained with Subaru Telescope equipped with Suprime-Cam mosaic camera. The wide-field ($20\arcmin \times 24\arcmin$) photometry catalog of 38,856 objects ($V \sim 16-26$\,mag) is presented. This survey is also intended to serve as ``a finding chart'' for future imaging and spectroscopic observation programs of Leo\,A.
\end{abstract}


\section{Introduction}
Dwarf irregular galaxies have been supposed to be the simplest of gas-rich stellar systems. Presumably complex Milky Way-size galaxies have accreted many of them as ``building blocks''. Since a significant part of this population was affected/consumed during mergers, finding samples of isolated and unperturbed dwarf irregular galaxies and studying their evolution through the Hubble time is challenging. Analysis of these pristine systems, which were likely built-up only via self-enrichment, could provide insight into the star formation history of ``building blocks''.

Leo\,A -- an apparently isolated dwarf irregular galaxy in the Local Group -- could serve as an example of such a ``building block''. It is an extremely gas-rich \citep{Young1996} dark matter dominated stellar system \citep{Brown2007} of low stellar mass \citep{Cole2007} and metallicity \citep{vanZee2006}. It consists of multiple stellar populations ranging from $\sim$10\,Myr to $\sim$10\,Gyr old \citep[][and references therein]{Cole2007}. The present-day star formation activity is traced by H\,II regions, while existence of old stellar population is proved by detection of RR\,Lyr stars \citep{Dolphin2002}.

Studies of stellar content in Leo\,A were performed with {\it Hubble Space Telescope} ({\it HST}) WFPC2 \citep{Tolstoy1998,Schulte-Ladbeck2002} and ACS \citep{Cole2007} by imaging of the central part. However, photometric mapping to a comparable depth of the whole extent of galaxy's outskirts \citep[Holmberg's dimension $\sim$$7\arcmin \times 4.6\arcmin$;][]{Mateo1998} requires a wide-field imaging capability. The Subaru Telescope, equipped with Suprime-Cam mosaic camera, was ideally suited for this task and employed for our survey.

Structural analysis of Leo\,A by \citet{Vansevicius2004}, based on the same Suprime-Cam image set, has revealed a two-component disk/halo-like structure of the radial stellar surface-density profile and a sharp stellar edge of Leo\,A. These findings are confirmed using the new photometry catalog, which is based on the improved reduction procedures. Here we present the stellar photometry catalog and discuss Suprime-Cam data reductions.

The following parameters of Leo\,A, as derived from the red giant branch (RGB) star distribution by \citet{Vansevicius2004}, are adopted in this study: (1) center coordinates of the galaxy, $\alpha_{\rm Leo\,A}=9^{\rm h}59^{\rm m}24.0^{\rm s}$, $\delta_{\rm Leo\,A}=+30\arcdeg 44\arcmin 47.0\arcsec$ (J2000); (2) ellipticity -- ratio of semi-minor to semi-major axis, $b/a = 0.6$; (3) position angle of the major axis, ${\rm P.A.} = 114$\arcdeg; (4) size of Leo\,A along the semi-major axis, $a = 8$\arcmin.

The foreground interstellar extinction towards Leo\,A of $E(B-V) = 0.021$, is derived from the extinction maps \citep{Schlegel1998}. The distance to Leo\,A is assumed to be of 800\,kpc \citep[$1\arcmin \equiv 230$\,pc;][]{Dolphin2002}.

\section{Observations and Data Reductions}
\subsection{Suprime-Cam Imaging}
Taking into account the angular size of Leo\,A \citep[$\sim$$16\arcmin$;][]{Vansevicius2004}, the Subaru Telescope, equipped with the Prime Focus Camera \citep[Suprime-Cam;][]{Miyazaki2002}, was used to study the stellar content at the galaxy's very outskirts. A single-shot Suprime-Cam mosaic ($5 \times 2$ CCD chips; $2{\rm K} \times 4{\rm K}$ each) covers a field of $34\arcmin \times 27\arcmin$ (scale $0.2\arcsec\,{\rm pixel}^{-1}$), and magnitude of $V \sim 25$\,mag is reached in 60\,s.

We have observed Leo\,A on two nights of November 2001. Images of Leo\,A were acquired through $B$, $V$, and $I_{\rm C}$ (tagged as ``$I$'' throughout) Johnson-Cousins system passbands in clear sky and good seeing conditions (full width at half maximum of stellar images, ${\rm FWHM} \sim 0.8\arcsec$); see Table~1 for observation details. The analyzed field was centered at $\alpha_{\rm SC}=9^{\rm h}59^{\rm m}26.0^{\rm s}$, $\delta_{\rm SC}=+30\arcdeg 46\arcmin 50.0\arcsec$ (J2000), i.e., slightly above the central part of Leo\,A to keep the main body of galaxy around the middle of a single CCD; see Figure~1 for the stacked mosaic color image of the survey field.

``Long'' and ``short'' exposures, consisting of five and three shots, respectively (see Table~1), were used to ensure a wide dynamical range ($V \sim 16-26$\,mag) of the final photometry catalog. The telescope was dithered between shots to avoid gaps midst CCDs while mapping the field. The applied dithering pattern of ``long'' (5 shots; $0\arcmin$, $0\arcmin$; $+1\arcmin$, $-2\arcmin$; $+2\arcmin$, $+1\arcmin$; $-1\arcmin$, $+2\arcmin$; $-2\arcmin$, $+1\arcmin$ in $\alpha$ \& $\delta$, respectively) and ``short'' (3 shots; $0\arcmin$, $0\arcmin$; $+0.33\arcmin$, $-1.67\arcmin$; $+1.33\arcmin$, $-1.2\arcmin$ in $\alpha$ \& $\delta$, respectively) exposures assured an internal calibration of astrometry and photometry, relying on the same stars measured in adjacent CCD chips.

The standard reduction procedures (bias subtraction and flat-fielding) were performed with the software package SDFRED\footnote{SDFRED: http://subarutelescope.org/Observing/Instruments/SCam/sdfred/sdfred1.html.en} \citep{Yagi2002,Ouchi2004}, dedicated to the Suprime-Cam data. For flat-fielding we used ``dark night sky'' flats, which were median combined from images of uncrowded sky regions of a M\,33 survey, conducted in parallel to Leo\,A observations. From the analyses of the sky background across images we estimated the overall flat-fielding error to be of $\lesssim$2\% for the $I$ passband and smaller for $B$ and $V$. Note, however, that our internal photometric calibration procedure involved linear corrections of $\lesssim$0.03\,mag photometric gradients at the edges of the mosaic derived from overlapping images, since ``dark night sky'' flats do not provide an exact representation of global sensitivity variation across wide-field CCD images \citep[see, e.g.,][]{Chromey1996}, however, they remove small-scale variations well.

The ``StarL'' program \citep{Narbutis2009} was used to perform corrections of image defects, e.g., ``blooming streaks'' of saturated stars, and produce a defect-free mosaic image shown in Figure~1.

\subsection{Coordinate System}
For object cross-identification between different frames, exposures, and passbands their coordinates, ($x,y$), have to be transformed to a reference coordinate system, ($X,Y$), of a surveyed field. Therefore, we adopted the SDFRED \citep{Yagi2002,Ouchi2004} software package to create a reference mosaic image and to find transformation equations for each CCD frame to the reference coordinate system, ($X,Y$).

The SDFRED automatically corrected image distortions according to the prescription from the optical ray-tracing of Suprime-Cam and differential atmospheric refraction dependence on airmass \citep{Miyazaki2002}. The accuracy of $\lesssim$1.5\,pix (0.3\arcsec) was achieved by this procedure between different passbands and exposures, but it was not suitable to cross-identify objects accurately enough for the photometric catalog.

Therefore, we stacked distortion-corrected images to make a reference $V$ passband mosaic with the pixel coordinate system, ($X,Y$). Using IRAF's \verb"geomatch", we computed coordinate transformation equations, $T(x,y)$, for each image by interactively fitting a 5$^{\rm th}$-order polynomial to a sample of points distributed uniformly over an image, i.e., $(x,y) \ast T(x,y) \rightarrow (X,Y)$. This procedure improved the accuracy of coordinate match between different passbands and exposures, leaving residuals of $\lesssim$0.4\,pix (0.08\arcsec).

The astrometric World Coordinate System (WCS) solution of the reference $V$ passband mosaic was derived by cross-correlating a list of reference objects with the USNO-B1.0 catalog of standard astrometric stars using IRAF's \verb"ccmap". Since the mosaic was transformed with the SDFRED to the tangential sky projection, only linear terms were needed. The fitting procedure was performed by iteratively removing outliers and resulted in $\sim$150 uniformly distributed stars used to get an astrometric solution with r.m.s. of $\sim$0.2\arcsec. Finally, the pixel coordinate system of the reference mosaic, ($X,Y$), was tied to the equatorial sky (J2000) coordinate system, ($\alpha,\delta$).

\subsection{PSF Photometry}
Due to variable point spread function (PSF; FWHM changes up to $\sim$25\% across the Suprime-Cam field, especially at mosaic edges) for analysis we used only six central CCDs ($20\arcmin \times 24\arcmin$) covering the whole extent of Leo\,A (see Figure~1). Since corrections of geometric image distortions (i.e., re-sampling to a tangential sky projection) and image stacking degrade the quality of PSF, stellar photometry was performed on the original flat-fielded CCD frames. Subsequently, aperture corrections, zero-point offsets, and residual gradients were applied via overlapping frames.

We have performed stellar PSF photometry on all individual ``long'' exposure frames by applying the DAOPHOT \citep{Stetson1987} implemented in the IRAF software package \citep{Tody1993}. We followed the photometry procedures described in the DAOPHOT documentation, however, in order to automate them and select stars for PSF modeling, we used custom-built IRAF CL scripts and external programs. The Gaussian PSF models as linear functions of image coordinates, ($x,y$), are based on reliable candidates of PSF stars (typical number $\sim$30), distributed uniformly across each CCD frame.

The PSF photometry was performed with IRAF's \verb"allstar". The residual images were searched for new objects uncovered after subtraction of bright companions by setting a higher detection threshold of $5\times\sigma_{\rm sky}$ (a detection threshold for the first iteration was set to $3\times\sigma_{\rm sky}$) to avoid spurious detections. Newly found objects were appended to the initial object list and the \verb"allstar" procedure was repeated. Tests have shown that two \verb"allstar" iterations were sufficient to detect stars even in the crowded central part of Leo\,A.

Finally, neighbors of PSF stars were removed from images by subtracting their best-fit models. We measured the ``total'' magnitudes, $m^{\rm total}$, of those PSF stars using the $1.75 \times {\rm FWHM}$ radius apertures. These magnitudes were used as references to transform the ``allstar'' magnitudes, $m^{\rm allstar}$, to the ``total'' magnitude system of the image. Due to PSF variation across the image, aperture corrections are linear functions of coordinates in CCD frames with typical r.m.s. of the fit $\lesssim$0.005\,mag, assuring that ``allstar'' magnitudes of stars are reliably transformed to the ``total'' magnitude system of the mosaic. We note, that residual aperture corrections from $m^{\rm total}$ to the true magnitudes, measured through large $3.5 \times {\rm FWHM}$ radius aperture, are constant for individual exposures on each CCD, and were taken into account as zero-point offsets.

Since the fields are relatively sparse and there are too few bright stars suitable for PSF modeling in ``short'' exposure frames, the aperture photometry was performed, and bright objects appended to the PSF photometry list.

\subsection{The Instrumental Photometric System}
Since the exposure dithering scheme was used, the same object was measured in several individual frames, in some cases obtained with different CCDs. This assured an accurate internal calibration of instrumental magnitudes over the survey field by using the same stars from overlapping frames.

For each passband we derived a global photometric solution of instrumental magnitudes over the whole survey field. Photometry lists of objects measured in individual CCD frames were merged by positional coincidence adopting the maximum allowed coordinate difference of 2\,pix (0.4\arcsec). The final instrumental magnitudes were computed by taking median values of the zero-point corrected magnitudes measured in individual exposures. Some stars in crowded regions were measured as ``double'' objects, i.e., artificially split by DAOPHOT. Those objects were analyzed interactively and considered to be single summing up their magnitudes, if distance between them was less than 1.5\,pix ($<$0.3\arcsec).

The initial ``long'' exposure object list was produced in three steps: 1) all objects in each passband detected in 2 or more individual exposures were selected ($B$ -- 62,095, $V$ -- 52,794, $I$ -- 57,533 objects); 2) the object lists in three passbands were merged by positional coincidence adopting the maximum allowed coordinate difference of 2\,pix (0.4\arcsec); 3) objects possessing $V$ and at least one of the $B$ or $I$ passband magnitudes were included in the object list.

``Short'' exposure aperture photometry was transformed to the ``long'' exposure system by applying zero-point corrections. The bright ($V \lesssim 20.5$\,mag) stars from ``short'' exposures (saturated in ``long'' exposures) were included in the final catalog. Finally, based on visual inspection of color mosaic images, the object list was interactively cleaned from obvious background galaxies to produce the instrumental magnitude catalog, consisting of 38,856 objects.

Note, however, that instrumental magnitudes were derived neglecting differences in color equations of individual CCDs, since the quantum efficiency curves of individual CCDs \citep{Miyazaki2002} differ only by $3-4$\% and make these corrections negligible.

\subsection{The Standard Photometric System}
To transform instrumental magnitudes to the standard photometric system we used stellar photometry of Leo\,A obtained by {\it HST} observations. The WFPC2 field observed by \citet{Tolstoy1998} was imaged through the F439W, F555W, and F814W passband filters. This data set was measured applying PSF photometry by \citet{Holtzman2006} and published on the ``{\it HST} Local Group Stellar Photometry Archive''\footnote{http://astronomy.nmsu.edu/holtz/archival/html/lg.html}, providing stellar magnitudes in $B$, $V$, and $I$ passbands of the standard Johnson-Cousins system, suitable for Suprime-Cam calibration.

The WFPC2 field is small ($\sim$$2.4\arcmin \times 2.4\arcmin$) and overlaps only with one of six Suprime-Cam CCDs used in our study. However, an accurate internal photometric calibration of individual Suprime-Cam CCDs, having a maximum magnitude difference of $\lesssim$0.02\,mag for bright stars ($V < 23$\,mag), enabled us to use local photometric standards \citep{Holtzman2006} to calibrate our instrumental magnitudes accross the entire field.

We have selected accurately measured ($V \lesssim 22.5$\,mag; ${\rm FLAG} = 0$) isolated (${\rm CROWD} = 0$) stars from the \citet{Holtzman2006} catalog and cross-identified them with reliably measured stars from our catalog, allowing a maximum coordinate miss-match of $\lesssim$0.2\arcsec.

This resulted in 161 photometric standard, spanning the $V$ magnitude range from $\sim$20.0 to $\sim$22.5\,mag and wide color ($B-V$, $V-I$) ranges from $\sim$$-0.3$ to $\sim$2.0\,mag. Figures~2(a) \& (b) show $V$ versus $V-I$ diagrams of stellar objects in the WFPC2 \citep[5,478 stars;][]{Holtzman2006} and Suprime-Cam (9,739 stars located inside the ellipse of $a < 3.5$\arcmin) fields.

The following calibration equations were derived to transform instrumental magnitudes to the standard photometric system:
\begin{eqnarray}
(B-V) = -0.051 + 1.142 \times (B-V)_{\rm instr}\,, \\
(V-I) = 0.412 + 1.013 \times (V-I)_{\rm instr}\,, \\
V = 2.490 + V_{\rm instr} + 0.015 \times (V-I)_{\rm instr}\,.
\end{eqnarray}

Figures~2(c) \& (d) show residuals of calibrated $V$ magnitude, $V_{\rm SC} - V_{HST}$, versus $B-V$ and $V-I$ colors, respectively. Residuals of colors, $(B-V)_{\rm SC} - (B-V)_{HST}$ and $(V-I)_{\rm SC} - (V-I)_{HST}$, are shown versus corresponding colors in Figures~2(e) \& (f), respectively. Photometric standards used for calibration (158 for $B-V$ and 161 for $V-I$) are indicated by open circles. The r.m.s. scatter of residuals is of $\sim$0.01\,mag. Taking into account the number of standard stars, a transformation accuracy of $\sim$0.003\,mag in all passbands was achieved.

\section{Results}
The Suprime-Cam photometry catalog of 38,856 objects ($V \sim 16-26$\,mag) in Leo\,A field ($20\arcmin \times 24\arcmin$) centered at $\alpha_{\rm SC}=9^{\rm h}59^{\rm m}26.0^{\rm s}$, $\delta_{\rm SC}=+30\arcdeg 46\arcmin 50.0\arcsec$ is presented in Table~2. Object coordinates, $V$ magnitude, $B-V$ and $V-I$ colors, corresponding photometric errors ($\sigma_{V}$, $\sigma_{B-V}$, and $\sigma_{V-I}$), the number of individual measurements in $V$ passband, $sharpness$ and $\chi^{2}$ parameters of the PSF model fit in $V$ passband, derived by DAOPHOT, are provided. There are 3,841 and 7,835 objects lacking $B-V$ and $V-I$ colors, respectively; 95 bright stars are measured in ``short'' exposure images. The fraction of objects having the following number of independent measurements is: 29\% -- 2, 24\% -- 3, 24\% -- 4, and 23\% -- 5. Although obvious background galaxies are not included in the catalog, 6,942 objects ($\sim$20\%) are classified and marked as extended based on PSF model fitting results and visual inspection of color mosaic images.

The parameters: $\sigma$, $sharpness$, and $\chi^{2}$ for 38,856 objects are plotted versus $B$, $V$, and $I$ magnitudes in Figure~3. Bright extended sources can be clearly distinguished in this parameter space, with the majority of them having $sharpness \gtrsim 0.5$ and $\chi^{2} \gtrsim 1.5$. Photometry results of stellar objects located inside the ellipse of semi-major axis $a < 5.5$\arcmin, i.e., the main body of Leo\,A, are displayed in $V$ vs. $B-V$ (14,831 stars) and $I$ vs. $V-I$ (12,154 stars) diagrams, shown in Figures~4(a) \& (b), respectively. The typical photometric errors, $\sigma$, are indicated with error-bars. The CMDs of Leo\,A show clear RGB population, main-sequence and blue-loop stars. The detection completeness of $\sim$50\% (see Table~1) is located just below the position of red clump stars in the CMDs. Although judging from Figures~2(a) \& (b), {\it HST} observations of Leo\,A central field by \citet{Tolstoy1998} have a comparable completeness limit to the Suprime-Cam data, we see a wider RGB in our data due to high field crowding. However, there is no strong crowding effect in the outer parts ($a \gtrsim 4$\arcmin) of the galaxy, where our survey was mainly targeted.

In order to estimate the completeness of the photometry catalogue in cases of various star densities more carefully, we performed artificial star tests (ASTs), following prescriptions by \citet{Stonkute2008} within a rectangular test area of $\sim$$13\arcmin \times 9\arcmin$ in size, centered on Leo A and extending beyond its limits \citep{Vansevicius2004}. We computed completeness\footnote{$N_{r}/N_{i} \times 100$\%, where $N_{i}$ and $N_{r}$ are the number of {\it input} and {\it recovered} artificial stars, respectively} for 9 $I$ passband reference magnitudes, evenly distributed in the range of $I\sim 20.5 - 25.5$\,mag. For every reference magnitude we produced two sets of AST frames. The original frames of individual exposures were populated with 8,000 AST stars, distributed on a regular grid with a step of $\sim$7\arcsec\ (i.e., $\sim$$10 \times {\rm FWHM}$ to avoid self-crowding) within the test area. The second set of AST frames was produced in the same way; however, coordinates of 8,000 AST stars were shifted in $\alpha$ and $\delta$ directions by $\sim$3.5\arcsec. This resulted in a total number of 16,000 AST spatial sampling points per reference magnitude.

To measure AST frames we used the same object detection and photometry procedures as for the original Leo\,A frames. For each of the 9 reference magnitudes photometric catalogs of all stars (containing a fraction of AST stars) detected in the AST test area were produced. We selected AST stars from these catalogs by matching their coordinates with the input AST star coordinates, allowing a maximum difference of 1.5\,pix (0.3\arcsec). Based on the results of AST star photometry we estimated the completeness of the photometry catalog for three crowding cases: inside an ellipse of $a < 2$\arcmin\ and within two elliptical rings of $\sim$0.6\arcmin\ in width at $a \sim 3$\arcmin\ and $a \sim 6$\arcmin\ radial distances. Completeness estimates as a function of the $I$ passband magnitude and radial distance from the galaxy center are shown in Figure~5. Brightest ($I < 22$\,mag) stars are confidently detected even in the crowded central part, however, at $I \sim 24$\,mag completeness already varies from $\sim$50\% to $\sim$85\% depending on the radial distance from the galaxy center.

\section{Discussion and Conclusions}
Since deep wide-field imaging became available, studies of dwarf galaxy structures have been a subject of great interest during the recent years \citep[see, e.g.,][and references therein]{Hidalgo2009}. Subaru Suprime-Cam has already been successfully used in analysis of other Local Group dwarf galaxies, e.g., the transitional galaxy DDO210 \citep{McConnachie2006}, Andromeda II dSph \citep{McConnachie2007}, and Ursa Major I dSph \citep{Okamoto2008}, among others.

We have performed a photometric survey of a field of $20\arcmin \times 24\arcmin$ size, covering the complete extent of an apparently isolated Local Group dwarf irregular galaxy Leo\,A, with Subaru Suprime-Cam in $B$, $V$, and $I$ passbands. In this study we present: (1) the photometry catalog of 38,856 objects ($V \sim 16-26$\,mag), about 80\% of which are classified as star-type; (2) $B$, $V$, and $I$ passband ``finding charts'' (stacked CCD mosaics), suitable for object identification, available via the WWW\footnote{http://www.astro.ff.vu.lt/data/leoa/}.

This photometry data set of Leo\,A has already been used to cross-identify sources in the {\it Spitzer} study on census of asymptotic giant branch stars by \citet{Boyer2009}. Therefore, the catalog and mosaic images could be useful for future imaging and spectroscopic observation programs. They complement the deepest available {\it HST}/ACS images of the central part of Leo\,A \citep[$I \sim 28$\,mag;][]{Cole2007}.

\acknowledgments
We are thankful to our collaborators: Chisato Ikuta, Pascale Jablonka, Kouji Ohta, Valdas Vansevi\v{c}ius, and Yoshihiko Yamada, whose help at the stage of observations and preliminary data reduction was invaluable. The Leo\,A survey is based on Suprime-Cam images, collected at the Subaru Telescope, which is operated by the National Astronomical Observatory of Japan. The research has made use of the following: the NASA/IPAC Extragalactic Database (NED), which is operated by the Jet Propulsion Laboratory, California Institute of Technology, under contract with the National Aeronautics and Space Administration; the SAOImage DS9, developed by Smithsonian Astrophysical Observatory; the USNOFS Image and Catalog Archive operated by the United States Naval Observatory, Flagstaff Station. The data presented in this paper were partly obtained from the Multimission Archive at the Space Telescope Science Institute. This research was partly funded by a grant (No. MIP-102/2011) from the Research Council of Lithuania.

\newpage
\begin{deluxetable}{c c c c} \tabletypesize{\scriptsize}
\setlength{\tabcolsep}{3.0pt} \tablewidth{0pc} \tablecolumns{4}
\tablecaption{Suprime-Cam Observations of Leo\,A}
\tablehead{
\colhead{Passband} &
\colhead{Exposures\tablenotemark{a}} &
\colhead{Seeing\tablenotemark{b}} &
\colhead{$m_{\rm 50\%}$\tablenotemark{c}} \\
\colhead{} &
\colhead{$n \times {\rm [s]}$} &
\colhead{[\arcsec]} &
\colhead{[mag]}}
\startdata
$B$ & $5 \times 600$ & 0.7 -- 0.9 & 26.2 \\
$B$ & $3 \times 20$ & 1.0 -- 1.4 & \nodata \\
$V$ & $5 \times 360$ & 0.6 -- 0.9 & 25.8 \\
$V$ & $3 \times 20$ & 0.8 -- 0.9 & \nodata \\
$I$ & $5 \times 240$ & 0.7 -- 0.8 & 24.8 \\
$I$ & $3 \times 20$ & 0.5 -- 0.6 & \nodata
\enddata
\tablenotetext{a}{~Number and duration (seconds) of exposures}
\tablenotetext{b}{~Range of stellar FWHM}
\tablenotetext{c}{~Limiting magnitude at detection completeness of 50\%}
\end{deluxetable}

\newpage
\begin{deluxetable}{c c c c c c c c c c c c c} \tabletypesize{\scriptsize}
\setlength{\tabcolsep}{3.0pt} \tablewidth{0pc} \tablecolumns{13}
\tablecaption{The Suprime-Cam Photometry Catalog in the Leo\,A Field}
\tablehead{
\colhead{ID} &
\colhead{$\alpha_{\rm J2000}$\tablenotemark{a}} &
\colhead{$\delta_{\rm J2000}$\tablenotemark{a}} &
\colhead{$V$\tablenotemark{b}} &
\colhead{$B-V$\tablenotemark{b}} &
\colhead{$V-I$\tablenotemark{b}} &
\colhead{$\sigma_{V}$\tablenotemark{c}} &
\colhead{$\sigma_{B-V}$\tablenotemark{c}} &
\colhead{$\sigma_{V-I}$\tablenotemark{c}} &
\colhead{$n$\tablenotemark{d}} &
\colhead{$sh$\tablenotemark{e}} &
\colhead{$\chi^{2}$\tablenotemark{e}} &
\colhead{Type\tablenotemark{f}}}
\startdata
J095839.45+305247.5 & 09:58:39.451 & +30:52:47.51 & 24.486 & 0.400 & 0.449 & 0.014 & 0.030 & 0.171 & 3 & 0.31 & 0.77 & 0 \\
J095839.45+305323.5 & 09:58:39.455 & +30:53:23.52 & 24.788 & 0.577 & 1.326 & 0.055 & 0.113 & 0.072 & 2 & 0.47 & 0.91 & 1 \\
J095839.46+305453.7 & 09:58:39.464 & +30:54:53.70 & 23.708 & 0.972 & 1.281 & 0.490 & 0.491 & 0.531 & 3 & 0.80 & 1.16 & 1 \\
J095839.47+305128.1 & 09:58:39.468 & +30:51:28.13 & 25.788 & 0.743 & 1.424 & 0.201 & 0.312 & 0.352 & 2 & -0.72 & 0.72 & 0 \\
J095839.51+305241.5 & 09:58:39.513 & +30:52:41.51 & 25.803 & 0.394 & \nodata & 0.124 & 0.175 & \nodata & 2 & -0.36 & 0.67 & 0
\enddata
\tablecomments{The photometry catalog of objects residing in the field of $20\arcmin \times 24\arcmin$, centered at $\alpha_{\rm SC}=9^{\rm h}59^{\rm m}26.0^{\rm s}$, $\delta_{\rm SC}=+30\arcdeg 46\arcmin 50.0\arcsec$. This table is available in its entirety in a machine-readable form in the online journal. A portion is shown here for guidance regarding its form and content.}
\tablenotetext{a}{~R.A. \& Decl. (J2000) coordinates in the USNO-B1.0 catalog system}
\tablenotetext{b}{The Johnson-Cousins system}
\tablenotetext{c}{~Photometric errors: $\sigma_{B-V} = (\sigma_{B}^{2}+\sigma_{V}^{2})^{1/2}$, $\sigma_{V-I} = (\sigma_{V}^{2}+\sigma_{I}^{2})^{1/2}$}
\tablenotetext{d}{~Number of measurements in $V$ passband; $n_{\rm max} = 5$}
\tablenotetext{e}{~$sharpness$ and $\chi^{2}$ -- DAOPHOT parameters of PSF model fitting in $V$ passband; not available for bright stars ($V \lesssim 20$\,mag) measured by aperture photometry in ``short'' exposure frames}
\tablenotetext{f}{~Classification based on PSF model fitting and visual inspection of images: 0 -- star, 1 -- extended object}
\end{deluxetable}

\newpage
\begin{figure}
\epsscale{0.8}
\plotone{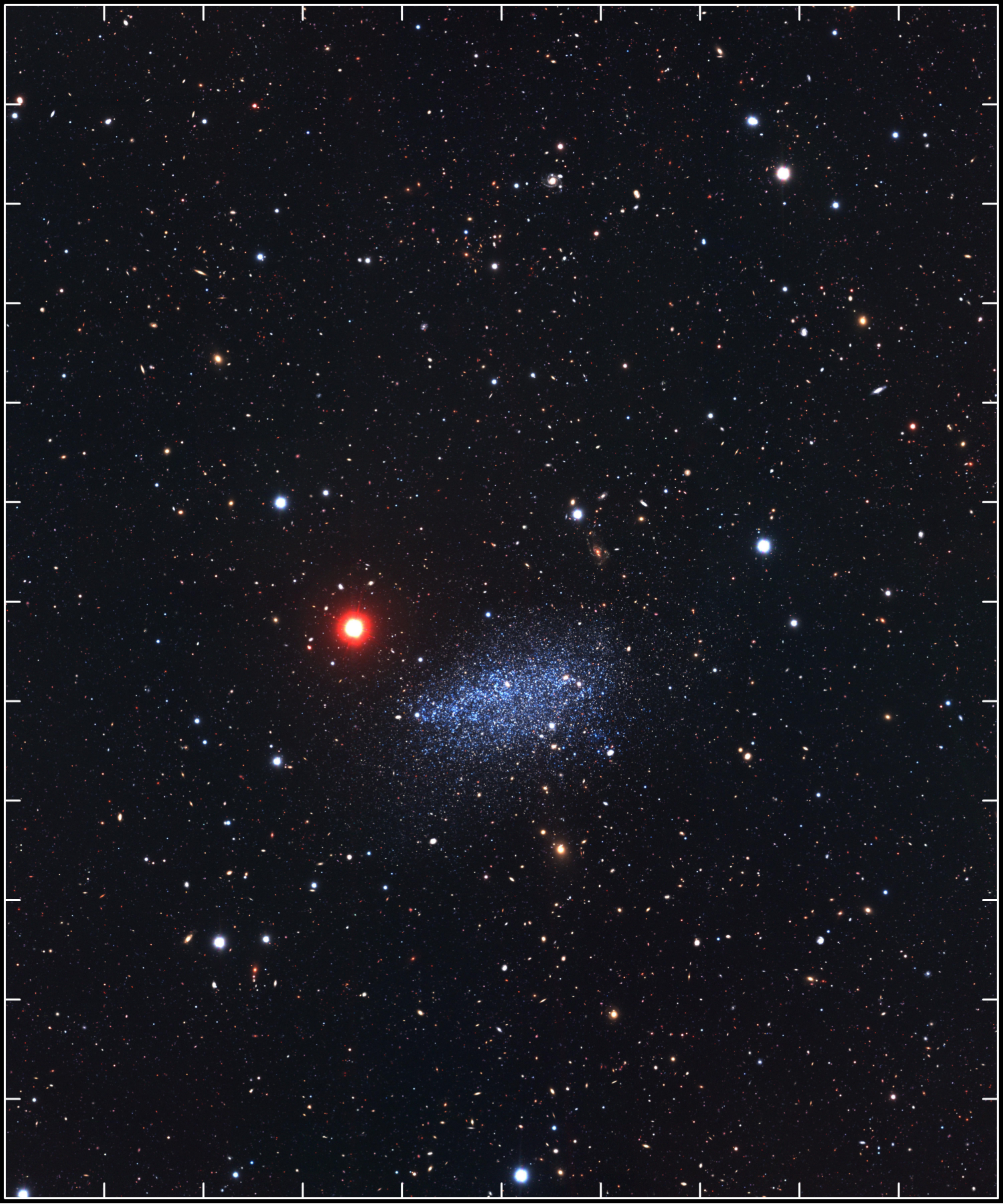}
\caption{Suprime-Cam image of the Leo\,A galaxy, displaying $20\arcmin \times 24\arcmin$ field (covered by the photometry catalog) in $B$ ({\it blue}), $V$ ({\it green}), and $I$ ({\it red}) passbands, centered at $\alpha_{\rm SC}=9^{\rm h}59^{\rm m}26.0^{\rm s}$, $\delta_{\rm SC}=+30\arcdeg 46\arcmin 50.0\arcsec$. North is up, east is left; tick-marks are placed at every 2\arcmin\ on ($\alpha,\delta$) axes.}
\end{figure}

\newpage
\begin{figure}
\epsscale{0.8}
\plotone{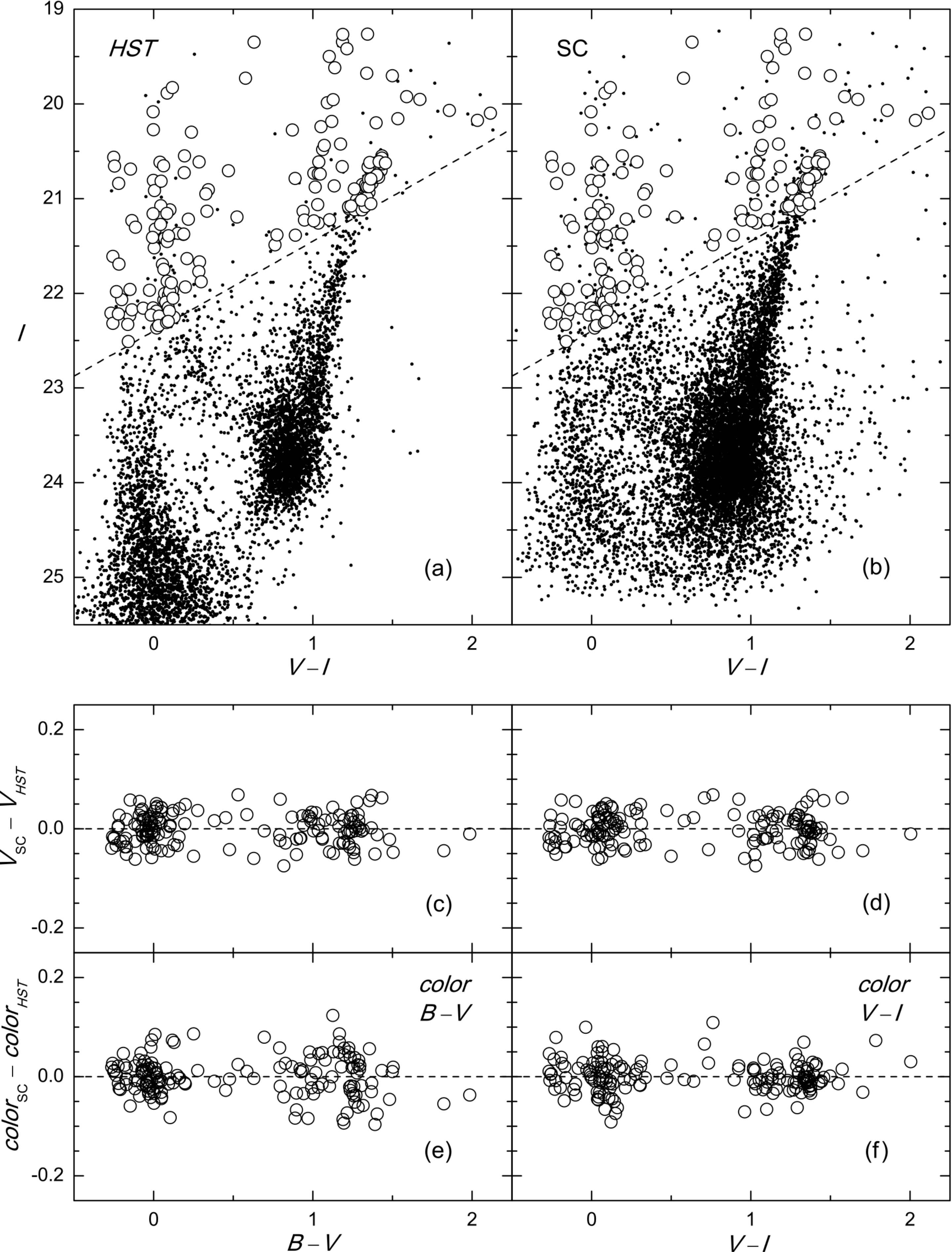}
\caption{Photometric calibrations of Leo\,A Suprime-Cam (SC) catalog via {\it HST} data. $V$ vs. $V-I$ diagrams of stellar objects in the WFPC2 field \citep[5,478 stars;][]{Holtzman2006} and Suprime-Cam (9,739 stars located inside the ellipse of $a < 3.5$\arcmin) are shown in (a) and (b), respectively; dashed line, $I \sim 22.5 - 0.95 \times (V-I)$, marks $V = 22.5$\,mag selection limit of the photometric standards. (c) and (d) show residuals of $V$ magnitude, $V_{\rm SC} - V_{HST}$, vs. $B-V$ and $V-I$ colors, respectively. Residuals of colors, $(B-V)_{\rm SC} - (B-V)_{HST}$ and $(V-I)_{\rm SC} - (V-I)_{HST}$, are shown vs. corresponding colors in (e) and (f), respectively. Photometric standards used for calibration (158 for $B-V$ and 161 for $V-I$) are indicated by open circles.}
\end{figure}

\newpage
\begin{figure}
\epsscale{0.9}
\plotone{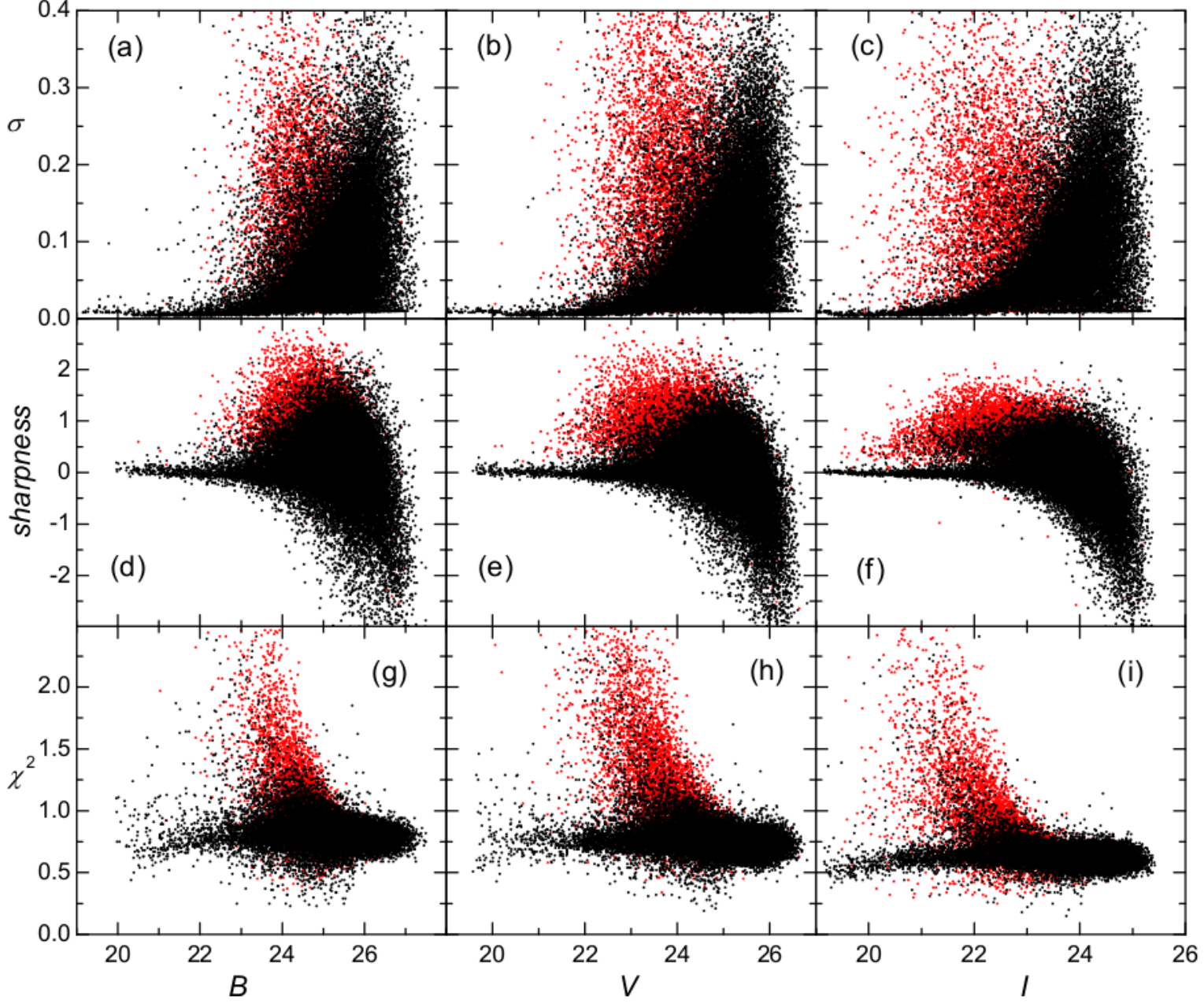}
\caption{Photometry parameters of objects in the Leo\,A field; stars and extended objects are indicated by {\it black} and {\it gray} ({\it red} in the online version) dots, respectively. Parameters are plotted vs. magnitude in $B$ (29,323 stars; 5,692 extended objects), $V$ (31,914; 6,942), and $I$ (24,921; 6,100) passbands: (a)--(c) photometric errors, $\sigma$; (d)--(f) object profile sharpness in respect to PSF, $sharpness$; (g)--(i) accuracy of object profile fitting with PSF, $\chi^{2}$. A color version of this figure is available in the online journal.}
\end{figure}

\newpage
\begin{figure}
\epsscale{0.95}
\plotone{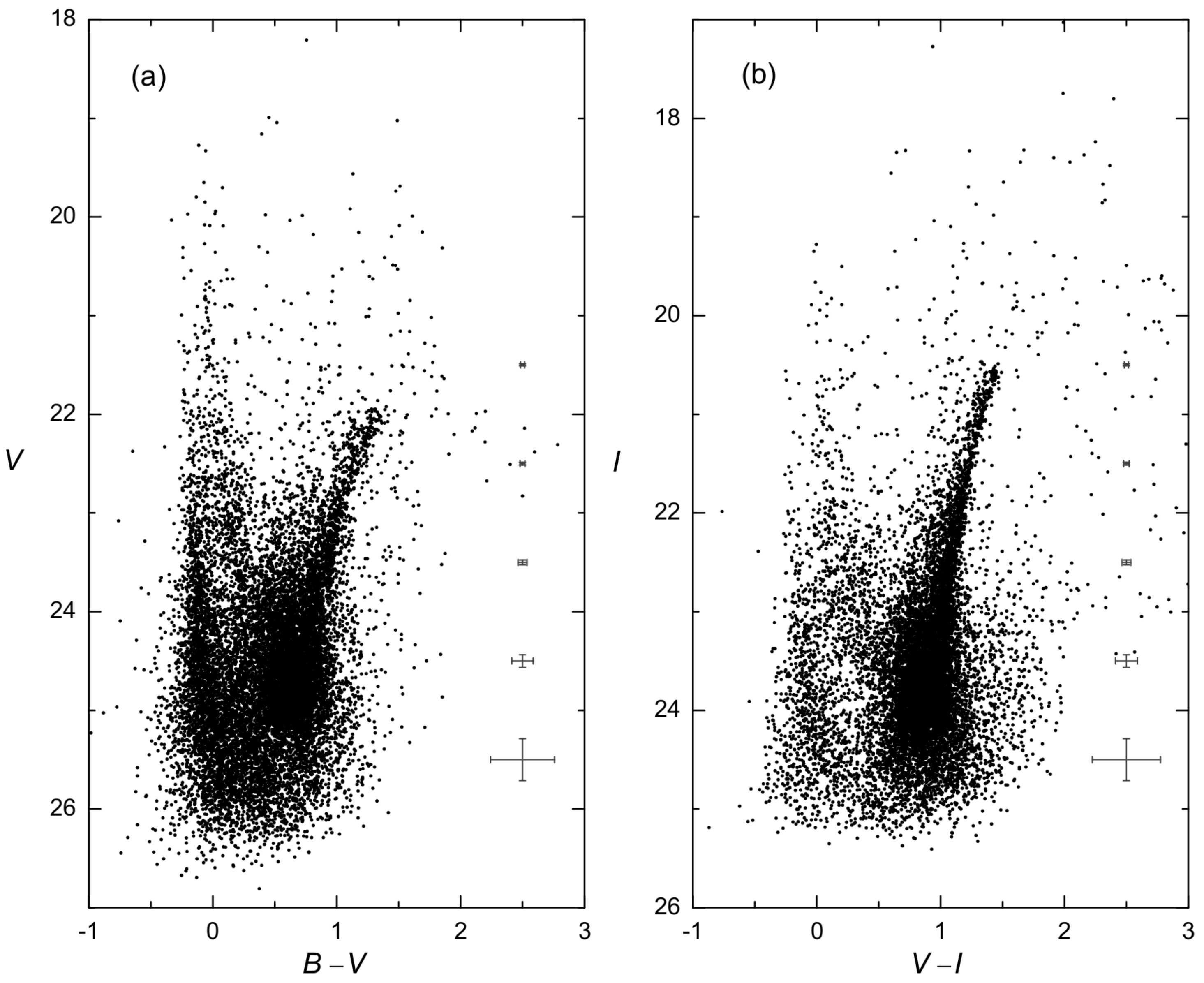}
\caption{Photometry results of star-type objects located inside the ellipse of $a < 5.5$\arcmin\ in the Leo\,A field. $V$ vs. $B-V$ (14,831 stars) and $I$ vs. $V-I$ (12,154 stars) diagrams are shown in (a) and (b), respectively. The typical photometric error, $\sigma$, is indicated with error-bars.}
\end{figure}

\newpage
\begin{figure}
\epsscale{0.5}
\plotone{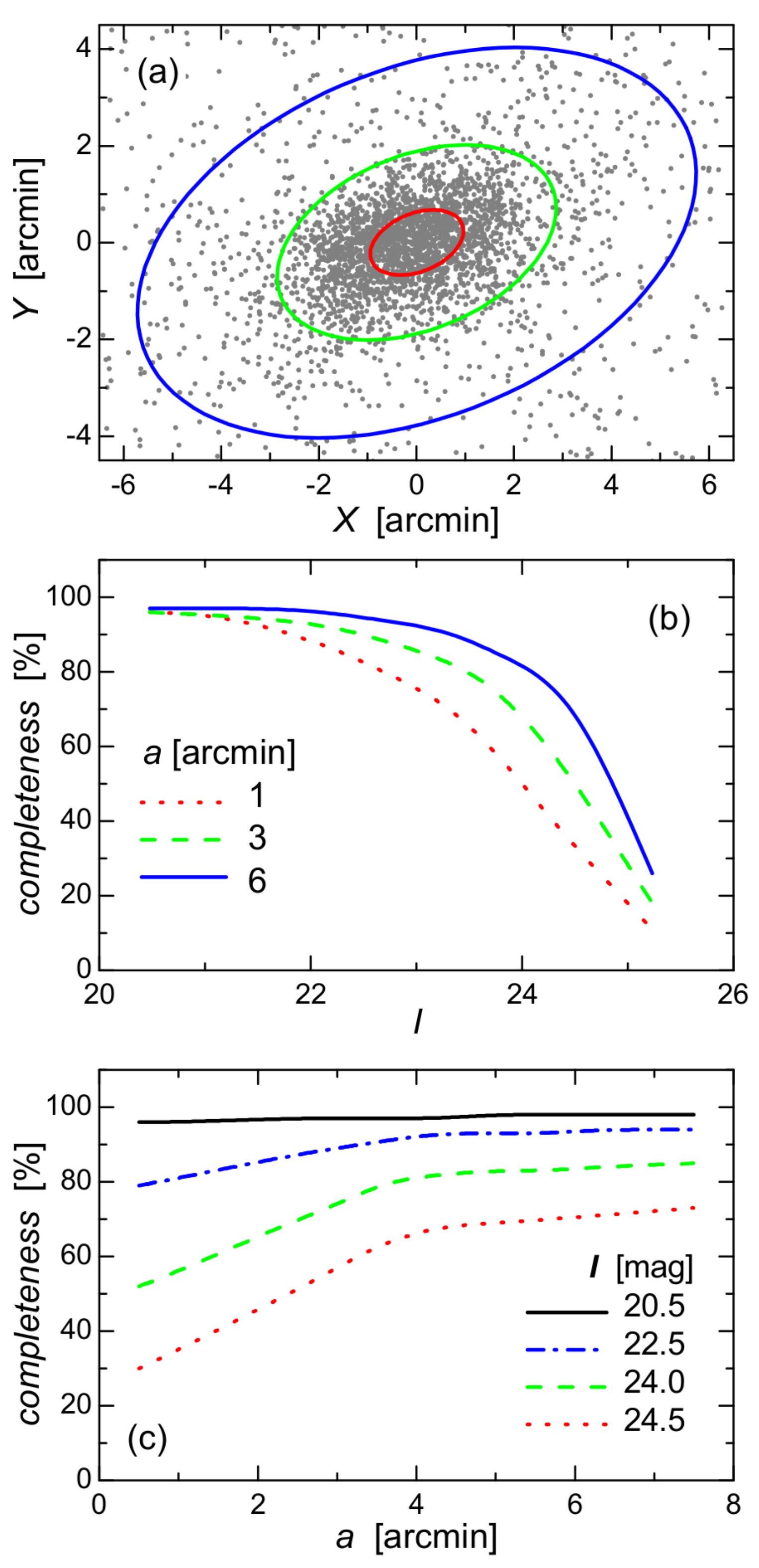}
\caption{Photometry completeness estimates based on artificial star tests of Suprime-Cam photometry in the Leo\,A field. Panel (a) -- the central part ($\sim$$13\arcmin \times 9\arcmin$) of the Leo\,A field; {\it gray} dots indicate 3,195 stars with $I \lesssim 23.0$\,mag; ellipses centered on Leo\,A of $a \sim 1\arcmin, 3\arcmin, 6\arcmin$ \citep[$b/a = 0.6$; ${\rm P.A.} = 114$\arcdeg;][]{Vansevicius2004} are overplotted. Panel (b) -- photometry completeness estimates for three radial distances from the galaxy center: inside the ellipse of $a < 2$\arcmin\ ({\it solid line}) and two elliptical rings of $\Delta a \sim 0.6$\arcmin\ width at $a \sim 3$\arcmin\ ({\it dashed}), and $\sim$$6\arcmin$ ({\it dotted}) plotted vs. $I$ magnitude. Panel (c) -- completeness estimates at four magnitudes: $I \sim 20.5$\,mag ({\it solid line}), $\sim$22.5\,mag ({\it dash-dotted}), $\sim$24.0\,mag ({\it dashed}), and $\sim$24.5\,mag ({\it dotted}) plotted vs. radial distance from the galaxy center along the semi-major axis, $a$. A color version of this figure is available in the online journal.}
\end{figure}

\end{document}